\documentclass[12pt]{article}
\usepackage{graphicx}
\catcode`\@=11
\def\numberbysection{\@addtoreset{equation}{section}
\def\theequation{\thesection.\arabic{equation}}}

\numberbysection
\newcommand{\abs}[1]{\left\vert#1\right\vert}
\newcommand{\be}{\[}
\newcommand{\beq}{\begin{equation}}
\newcommand{\bea}{\begin{eqnarray*}}
\newcommand{\beqa}{\begin{eqnarray}}
\renewcommand{\d}{{{\rm d}}}
\newcommand{\ds}{\displaystyle}
\newcommand{\de}{^{(2)}}
\newcommand{\dpar}{\partial}
\newcommand{\e}{{\rm e}}
\newcommand{\eq}{{\rm eq}}
\newcommand{\ee}{\]}
\newcommand{\eeq}{\end{equation}}
\newcommand{\eea}{\end{eqnarray*}}
\newcommand{\eeqa}{\end{eqnarray}}
\newcommand{\frad}[2]{\displaystyle{\displaystyle#1\over\displaystyle#2}}
\newcommand{\fd}{fluc\-tu\-a\-tion-dis\-si\-pa\-tion }
\newcommand{\g}{\gamma}
\newcommand{\h}{h}
\newcommand{\mean}[1]{\left\langle#1\right\rangle}
\renewcommand{\i}{{\rm i}}
\renewcommand{\l}{\ell}
\newcommand{\p}{\varphi}
\newcommand{\pp}{\pi}
\newcommand{\re}{\mathop{\rm Re}}
\newcommand{\Var}{\mathop{\rm Var}}
\newcommand{\un}{^{(1)}}
\newcommand{\unde}{^{(1,2)}}
\newcommand{\vb}{{\vphantom{M}}}

\newcommand{\D}{{\cal D}}

\renewcommand{\H}{{\cal H}}
\newcommand{\J}{J_{(\beta+1)/2}}
\renewcommand{\L}{L^{(\beta+1)/2}}
\renewcommand{\P}{{\cal P}}
\newcommand{\X}{{\cal X}}
\newcommand{\text}[1]{\textrm{#1}}

\def\binom#1#2{{#1\choose #2}}

\begin{document}
\centerline{\Large\bf Nonequilibrium dynamics of the zeta urn model}
\vspace{1cm}
\centerline{\large
by C.~Godr\`eche$^{a,}$\footnote{godreche@spec.saclay.cea.fr}
and J.M.~Luck$^{b,}$\footnote{luck@spht.saclay.cea.fr}}
\vspace{1cm}
\centerline{$^a$Service de Physique de l'\'Etat Condens\'e,
CEA Saclay, 91191 Gif-sur-Yvette cedex, France}
\vspace{.1cm}
\centerline{$^b$Service de Physique Th\'eorique\footnote{URA 2306 of CNRS},
CEA Saclay, 91191 Gif-sur-Yvette cedex, France}
\vspace{1cm}
\begin{abstract}
We consider a mean-field dynamical urn model, defined by rules
which give the rate at which
a ball is drawn from an urn and put in another one,
chosen amongst an assembly.
At equilibrium, this model possesses
a fluid and a condensed phase, separated by a critical line.
We present an analytical study of the nonequilibrium properties of
the fluctuating number of balls in a given urn,
considering successively the temporal evolution of its distribution,
of its two-time correlation and response functions,
and of the associated \fd ratio,
both along the critical line and in the condensed phase.
For well separated times the \fd ratio admits non-trivial limit values,
both at criticality and in the condensed phase,
which are universal quantities depending continuously on temperature.
\end{abstract}
\vfill
\vskip -3pt
\noindent P.A.C.S.: 02.50.Ey, 05.40.+j, 61.43.Fs.
\newpage
\section{Background: on dynamical urn models}

Dynamical urn models are simplified models of physical reality, which have
always played an important role in the elucidation of conceptual problems of
statistical mechanics and probability theory.

The ancestor and prototype of this class of models is the
Ehrenfest urn model~\cite{ehr}.
It was devised by P.~and T.~Ehrenfest, in their attempt to critically review
Boltzmann's $H-$theorem.
Consider $N$ balls, labeled from 1 to $N$, which
are distributed in two urns (or boxes).
At each time step a ball is chosen
at random (i.e., an integer between 1 and~$N$ is chosen at random), and
moved from the box in which it is, to the other box.
Let $N_{1}$ (respectively, $N_{2}$) be the numbers of balls in box number 1
(respectively, number~2) ($N_{1}+N_{2}=N$).
If the process is repeated
indefinitely, for any initial condition the system will relax to
equilibrium, characterized by a binomial distribution of balls in, say, box
number 1:
\begin{equation}
f_{k,\eq}=\P(N_{1}=k)=\binom{N}{k}\frac{1}{2^{N}}.
\label{fk2}
\end{equation}
This result is both intuitively translucent, and easy to derive (see below).
The partition function of the system is equal to (for $2$ boxes and $N$ balls)
\beq
Z(2,N)=\frac{2^{N}}{N!}.
\label{z2n}
\eeq
There are $2^N$ possibilities of distributing the balls amongst the 2 boxes,
however all $N!$ labelings of the balls are equivalent.

Finding the distribution of balls after $n$ steps,
$f_{k}(n)=\P(N_{1}(n)=k)$, requires more effort.
Kohlrausch and Schr\"{o}dinger~\cite{ks} found
the master equation for $f_{k}(n)$, which they interpreted as the
probability distribution of the position at time $n$ of a random walker
(played here by $N_{1}(n)$).
The full
solution of this master equation was given later on by Kac, Siegert and Hess
\cite{kac,kac2,sieg,hess}.

The Ehrenfest model is at the origin of a whole class of dynamical urn models,
which we name the \emph{Ehrenfest class}.
They generalize the original
Ehrenfest model in two ways, which we detail successively.

The first generalization consists in considering $M$ boxes instead of two.
Then, at equilibrium, the joint distribution of the occupation numbers
$N_{1}$, $N_{2}$, $\dots$, $N_{M}$, with
\[
\sum_{i=1}^{M}N_{i}=N,
\]
is the multinomial distribution
\[
\P(N_{1}=k_{1},\dots,N_{M}=k_{M})=\binom{N}{k_{1}\dots k_{M}}
\frac{1}{M^{N}},
\]
as a simple reasoning shows.
The marginal distribution of $N_{1}$ is
obtained by summation upon the other variables, and reads
\[
f_{k,\eq}=\P(N_{1}=k)=\binom{N}{k}\frac{1}{M^{N}},
\]
which is a simple generalization of~(\ref{fk2}).
In the thermodynamic
limit $N\rightarrow\infty $, $M\rightarrow\infty $, with fixed density
$\rho=N/M$, this yields a Poisson law:
\[
f_{k,\eq}=\e^{-\rho }\frac{\rho ^{k}}{k!}.
\]
The partition function of the system is equal to (for $M$ boxes and $N$
balls)
\begin{equation}
Z(M,N)=\frac{M^{N}}{N!},
\label{zinfty}
\end{equation}
as a simple generalization of the reasoning leading to~(\ref{z2n}) shows.

The dynamics of the Ehrenfest model and of its generalization to $M$ boxes
takes
place at infinite temperature because there is no constraint on the move of
the drawn ball.
The second direction of generalization consists in defining
these models at finite temperature, by introducing energy.
We assume that the energy is a sum of contributions of independent boxes:
\[
E(N_{1},\dots,N_{M})=\sum_{i=1}^{M}E(N_{i}),
\]
and choose a rule obeying detailed balance for the move of the drawn ball.
For instance, for the Metropolis rule, the move is
allowed with probability $\min(1,\exp(-\beta\Delta E))$.
Heat-bath dynamics is another possible choice (see below).

The fundamental ingredients for the definition of the models belonging to
the Ehrenfest class are therefore

\begin{description}
\item [(i)] the statistics: a ball is chosen at random, and put in a box
chosen at random,

\item [(ii)] the choice of
the energy of a box $E(N_{i})$, and of a dynamical rule (Metropolis
or heat-bath),

\item [(iii)] the geometry: for instance, boxes may be ordered on a line, or
on the contrary be all connected.
For short, we designate the latter geometry as the mean-field case.
\end{description}

The backgammon model~\cite{ritort} is a representative of the Ehrenfest class,
corresponding to the choice (where $\delta$ is Kronecker symbol)
\begin{equation}
E(N_{i})=-\delta(N_{i},0).\label{back}
\end{equation}
This model has been extensively studied, mainly in its mean-field
formulation~\cite{fr,gbm,gl}.

The dynamical (and equilibrium) properties of the Ehrenfest class depend
crucially on the choice of statistics described in (i), which we will
briefly refer to as the \emph{ball-box statistics}.
However, other choices
are possible, which define new classes of dynamical urn models.
The class which we will refer to for short as the \emph{Monkey class},
because it corresponds to the image of a monkey playing at exchanging
balls between boxes, is defined by:

\begin{description}
\item [(i)] the statistics: a box is chosen at random, from which any ball is
drawn, and put in another box, chosen at random (\emph{box-box statistics}),

\item [(ii)] the choice of energy and dynamical rule (as above),

\item [(iii)] the geometry (as above).
\end{description}

A first example of a model belonging to this class corresponds to taking
definition~(\ref{back}) for the energy~\cite{gbm}.
This model, referred to as model B
in~\cite{gbm}, possesses non-trivial dynamical properties~\cite{gbm,gl}.

A second example, inspired from quantum gravity, corresponds to
taking~\cite{bia}
\begin{equation}
E(N_{i})=\ln(N_{i}+1).
\label{defE}
\end{equation}
It presents interesting properties both at equilibrium~\cite{bia} and in
nonequilibrium situations~\cite{dgc}.
In contrast with the backgammon model, or with
model B, it possesses a phase transition between a fluid phase and a
condensed phase at finite temperature~\cite{bia}.
We will refer to this model as the \emph{zeta urn model}, for reasons which
will appear clear in the sequel.
The present work is entirely devoted
to the study of the nonequilibrium behavior of the zeta urn model
in the mean-field geometry.

Before specializing to this model, let us first present, in parallel,
some formalism which applies to the two classes of models defined above,
in order to underline the fundamental role played by the choice of
statistics for both the equilibrium and nonequilibrium properties of the
models.
Note that the equilibrium properties of the dynamical urn models
defined above are independent of the geometry, because boxes are independent.

For the Ehrenfest class, the partition function reads
\begin{equation}
Z(M,N)=\sum_{N_{1}}\cdots\sum_{N_{M}}\frac{\,p_{N_{1}}}{N_{1}!}\cdots
\frac{p_{N_{M}}}{N_{M}!}\;\delta\left(\sum_{i}N_{i},N\right),
\label{jstar}
\end{equation}
where
$$
p_{N_{i}}=\e^{-\beta E(N_i)}
$$
is the unnormalized Boltzmann weight attached to box number $i$.
For the Monkey class we have
\beq
Z(M,N)=\sum_{N_{1}}\cdots\sum_{N_{M}}\,p_{N_{1}}\cdots p_{N_{M}}\;\delta
\left(\sum_{i}N_{i},N\right).
\label{Zbb}
\eeq

Using the integral representation $2\i\pi\delta(m,n)=\oint\d z\,z^{m-n-1}$,
we obtain
\begin{equation}
Z(M,N)=\oint\frac{\d z}{2\i\pi z^{N+1}}\left[P(z)\right]^{M},
\label{contour}
\end{equation}
where
\begin{eqnarray}
P(z)=\sum_{k=0}^\infty\frac{p_{k}}{k!}\,z^k{\hskip.75cm}
&&(\mathrm{Ehrenfest}),
\label{PzE}\\
=\sum_{k=0}^\infty p_{k}\,z^{k}\qquad&&(\mathrm{Monkey}).
\label{Pzbb}
\end{eqnarray}
The equilibrium properties of the models are therefore entirely encoded in
the tem\-pe\-ra\-ture-de\-pen\-dent generating series $P(z)$.
The presence or absence of the factorial term $k!$
has direct implication on the analytic structure of the series, and by
consequence on the possible existence of a phase transition at finite
temperature.

The equilibrium probability distribution of the occupation number
$N_{1}$ reads, for the Ehrenfest class,
\begin{eqnarray}
f_{k,\eq} &=&\P(N_{1}=k)=\left\langle\delta\left( N_{1},k\right)
\right\rangle
\nonumber\\
&=&\frac{1}{Z(M,N)}\sum_{N_{1}}\dots\sum_{N_{M}}\delta\left(
N_{1},k\right)\frac{\,p_{N_{1}}}{N_{1}!}\dots\frac{p_{N_{M}}}{N_{M}!}
\;\delta\left(\sum_{i}N_{i},N\right)
\nonumber\\
&=&\frac{p_{k}}{k!}\frac{Z(M-1,N-k)}{Z(M,N)}.
\label{fkE}
\end{eqnarray}
For the Monkey class one obtains
\begin{equation}
f_{k,\eq}=p_{k}\frac{Z(M-1,N-k)}{Z(M,N)}.
\label{fkbb}
\end{equation}
At infinite temperature, equation~(\ref{contour}),
together with~(\ref{PzE}) and~(\ref{Pzbb}),
respectively lead to~(\ref{zinfty}) and to
\begin{equation}
Z(M,N)=\frac{(M+N-1)!}{(M-1)!\,N!}.
\label{jmonkey}
\end{equation}

In the thermodynamic limit at fixed density $\rho$,
the free energy per box is defined as
\[
\beta F=-\lim_{M\to\infty}\frac{1}{M}\ln Z(M,N),\qquad N\approx M\rho.
\]
At infinite temperature, equations~(\ref{zinfty}) and~(\ref{jmonkey}) yield
\bea
\lim_{\beta\to0}\,\beta F
=\rho\ln\rho-\rho{\hskip 3.525cm}&&(\mathrm{Ehrenfest}),\\
=\rho\ln\rho-(\rho+1)\ln(\rho+1)\qquad&&(\mathrm{Monkey}).
\eea
At finite temperature, the free energy can be obtained by evaluating the
contour integral in~(\ref{contour}) by the saddle-point method.
The saddle-point value $z_{s}$ of $z$
is a function of temperature and density through the equation
\beq
\frac{z_{s}P^{\prime }(z_{s})}{P(z_{s})}=\rho,
\label{col}
\eeq
and the free energy per box reads
\beq
\beta F=\rho\ln z_{s}-\ln P(z_{s}).
\label{free}
\eeq
Similarly, we obtain the following expressions
for the equilibrium occupation probabilities in the thermodynamic limit
\begin{eqnarray}
f_{k,\eq}=\frac{p_{k}}{k!}\frac{z_{s}^{k}}{P(z_{s})}\qquad
&&(\mathrm{Ehrenfest}),\label{fkeq1}\\
=p_{k}\frac{z_{s}^{k}}{P(z_{s})}{\hskip.91cm}&&(\mathrm{Monkey}).
\label{fkeq2}
\end{eqnarray}
This formalism will be illustrated below on the zeta urn model,
studied in this work.

For both classes of models, the temporal evolution of the
occupation probability
\[
f_{k}(t)=\P(N_{1}(t)=k)
\]
is given by the master equation

\begin{equation}
\frac{\d f_{k}(t)}{\d t}=\sum_{\l=0}^\infty
\left(\pp_{k+1,\l}+\pp_{\l,k-1}-\pp_{k,\l}-\pp_{\l,k}\right),
\label{master1}
\end{equation}
where $\pp_{k,\l}$ denotes the contribution of a move from the departure box,
containing $k$ balls, to the arrival box, containing $\l$ balls.
Restricting our study to the mean-field case,
we have (for $k,\l\geq 0$)
\begin{eqnarray}
\pp_{k,\l}=kf_{k}f_{\l}W_{k,\l}(1-\delta_{k,0})\qquad&&(\mathrm{Ehrenfest}),
\nonumber\\
=f_{k}f_{\l}W_{k,\l}(1-\delta_{k,0}){\hskip 1.05cm}&&(\mathrm{Monkey}),
\label{pkl}
\end{eqnarray}
where the term $1-\delta_{k,0}$ accounts for the fact that
the departure box cannot be empty,
and where the acceptance rate $W_{k,\l}$ depends on the dynamics chosen.
With the Metropolis rule we have
\[
W_{k,\l}=\min\left(1,\frac{p_{k-1}p_\l}{p_kp_{\l+1}}\right),
\]
while the heat-bath rule leads to~\cite{dgc}
\beq
W_{k,\l}=\frac{p_{\l+1}}{p_{\l}}
\left(\sum_{\l=0}^{\infty}f_{\l}\frac{p_{\l+1}}{p_{\l}}\right)^{-1},
\label{heat}
\eeq
which only depends on the label $\l$ of the arrival box.

In all cases, equation~(\ref{master1}) can be seen as the master equation of
a random walk for $N_{1}$, i.e., over the positive integers $k=0,1,\dots$,
\begin{eqnarray}
\frac{\d f_{k}(t)}{\d t} &=&\mu_{k+1}\,f_{k+1}+\lambda_{k-1}\,f_{k-1}-\left(
\mu_{k}+\lambda_{k}\right) f_{k}\qquad(k\ge1),
\label{master2a}\\
\frac{\d f_{0}(t)}{\d t} &=&\mu_{1}\,f_{1}-\lambda_{0}f_{0},
\label{master2b}
\end{eqnarray}
generalizing the result of Kohlrausch and
Schr\"{o}dinger for the Ehrenfest model.
In these equations,
\[
\lambda_k=\sum_{\l=0}^\infty\frac{\pp_{\l,k}}{f_k},\qquad
\mu_k=\sum_{\l=0}^\infty\frac{\pp_{k,\l}}{f_k}
\]
are, respectively, the hopping rate to the right,
corresponding to $N_{1}=k\rightarrow N_{1}=k+1$,
and to the left, corresponding to $N_{1}=k\rightarrow N_{1}=k-1$.
The equation for $f_{0}$ is special because one cannot select an empty box
as a departure box, nor can $N_1$ be negative, hence $\lambda_{-1}=\mu_0=0$.
In other words a partially absorbing barrier is present at site $k=0$.
The random walk is locally biased, to the right or to the left, according to
whether its velocity $\lambda_k-\mu_k$ is positive or negative, respectively.
It is easy to verify that~(\ref{master2a}) and~(\ref{master2b})
preserve the sum rules
\begin{eqnarray}
\sum_{k}f_{k}(t) &=&1,\label{sum1}\\
\sum_{k}k\,f_{k}(t) &=&\left\langle N_{1}(t)\right\rangle =\rho
\label{sum2},
\end{eqnarray}
expressing respectively the conservation of probability
and of the number of particles.

The equilibrium occupation probabilities~(\ref{fkeq1}),~(\ref{fkeq2})
are recovered as the unique stationary state
($\d f_k/\d t=0$) of the master equations~(\ref{master2a}),~(\ref{master2b}).
In contrast, except at infinite temperature, where $W_{k,\l}=1$,
so that the rates $\lambda_{k}$ and $\mu_{k}$ simplify,
the master equations cannot be solved explicitly.
The difficulty comes from the fact that
the rates $\lambda_{k}$ and $\mu_{k}$ are functions of the $f_{k}$,
hence the master equations are non-linear.
However, as will be illustrated by the present work,
the long-time behavior of these equations is amenable to analytic computations.

To summarize, for the two classes of dynamical urn models described above,
finding the properties of their equilibrium states is in general easy, but
attaining dynamical properties is much more difficult to achieve.
However if
one restricts the study to the fluctuating number of balls in a given box,
denoted by $N_{1}(t)$ all throughout this paper, it is,
in some cases, possible to predict
the long-time behavior of its probability distribution, $f_{k}(t)$, and also
of its two-time correlation and response functions.
This has been done in
previous studies for the backgammon model~\cite{fr,gbm,gl}.

In the present work we focus our interest on the mean-field
dynamical urn model
defined by the choice of energy~(\ref{defE}), and box-box statistics
--the zeta urn model.
We pursue the investigation of its
nonequilibrium properties, initiated in~\cite{dgc}.
The next section is devoted to a
more complete presentation of the model and to an outline of this paper.

\section{The zeta urn model}

At equilibrium, the zeta urn model is defined
by its partition function~(\ref{Zbb}), where the Boltzmann weight
\beq
p_{N_i}=\e^{-\beta E(N_i)}=(N_i+1)^{-\beta}
\label{boltz}
\eeq
corresponds to the choice of energy~(\ref{defE})~\cite{bia}.
The equilibrium phase diagram of the model~\cite{bia}
easily follows from the analysis of the previous section
(see~(\ref{contour})--(\ref{fkeq2})).

At low enough temperature $(\beta>2)$, there is a finite critical density:
\be
\rho_c=\frac{P'(1)}{P(1)}
=\frac{\zeta(\beta-1)-\zeta(\beta)}{\zeta(\beta)},
\ee
where $\zeta$ denotes Riemann's zeta function.

In the fluid phase $(\rho<\rho_c)$,
the equilibrium distribution~(\ref{fkeq2}) decays exponentially, since $z_s<1$.

At the critical density $(\rho=\rho_c)$, we have $z_s=1$, hence
\beq
f_{k,\eq}=\frac{p_k}{P(1)}=\frac{(k+1)^{-\beta}}{\zeta(\beta)},
\label{fkc}
\eeq
which is known as the zeta distribution.
In the regular part of the critical line~($\beta>3$),
the mean squared population is finite, and equal to
\be
\mean{N_i^2}=\sum_{k=0}^\infty k^2\,f_{k,\eq}
=\mu_c=\frac{\zeta(\beta-2)-2\zeta(\beta-1)+\zeta(\beta)}{\zeta(\beta)},
\ee
while it is infinite in the strong-fluctuation case~($2<\beta<3$).
Throughout the following, we shall restrict the study
to the regular part of the critical line.

In the condensed phase $(\rho>\rho_c)$,
a macroscopic condensate of particles appears.
Indeed, equation~(\ref{fkc}) still applies to all the boxes but one,
in which an extensive number of particles,
of order $N-M\rho_c=M(\rho-\rho_c)$, is condensed.

The dynamical definition of the model was given in the previous section.
For heat-bath dynamics, equation~(\ref{master1}),
together with~(\ref{pkl}) and~(\ref{heat}), leads to~\cite{dgc}
\beqa
&&\frad{\d f_k(t)}{\d t}=f_{k+1}(t)+\sigma(t)r_{k-1}f_{k-1}(t)
-\left(1+\sigma(t)r_k\right)f_k(t)\qquad(k\ge1),
\nonumber\\
&&\frad{\d f_0(t)}{\d t}=f_1(t)-\sigma(t)r_0f_0(t),
\label{df}
\eeqa
with
\be
r_k=\frac{p_{k+1}}{p_k}=\frac{f_{k+1,\eq}}{f_{k,\eq}}
=\left(\frac{k+1}{k+2}\right)^\beta
\ee
and
\be
\sigma(t)=\frad{1-f_0(t)}{\sum_{k=0}^\infty r_kf_k(t)}.
\ee
We assume that at time $t=0$ the system is quenched
from its infinite-temperature equilibrium state
to a finite temperature $T=1/\beta$.
Hence, by~(\ref{col}) and~(\ref{fkeq2}),
the initial occupation probabilities read
\beq
f_k(0)=\frac{\rho^k}{(1+\rho)^{k+1}}.
\label{fk0}
\eeq

In the fluid phase~($\rho<\rho_c$),
the equilibrium distribution $f_{k,\eq}$~(\ref{fkeq2})
is a stationary solution of~(\ref{df}), corresponding to $\sigma_\eq=z_s$.
The convergence of $f_k(t)$
towards $f_{k,\eq}$ is characterized by a finite relaxation time,
depending on $\beta$ and $\rho$.

The long-time behavior of the distribution $f_k(t)$,
both at criticality~($\rho=\rho_c$) and in the condensed phase~($\rho>\rho_c$),
is the subject of the next section.
In section 4 we establish the dynamical equations obeyed by the
two-time correlation and response functions of the population of a given box.
It is indeed well-known that
nonequilibrium properties are more fully revealed by
two-time observables~\cite{ck,aging}.
Section 5 is devoted to the analysis of the equilibrium properties of
these functions at criticality.
Section 6 is devoted to the analysis of their nonequilibrium properties at
criticality, including the violation of the fluctuation-dissipation theorem.
In section 7 we briefly investigate the nonequilibrium properties of the model
in the condensed phase.

\section{Long-time behavior of occupation probabilities}

\subsection{At criticality~($\rho=\rho_c$)}

We investigate how, starting from the disordered initial condition~(\ref{fk0}),
the occupation probabilities $f_k(t)$ converge toward
their critical equilibrium values $f_{k,\eq}$, given by~(\ref{fkc}),
which are the stationary solutions of equations~(\ref{df})
corresponding to $\sigma_\eq=1$.
In analogy with the analysis done in~\cite{dgc},
we anticipate that $\sigma(t)$ converges to this value as a power law:
\be
\sigma(t)\approx 1+A\,t^{-\omega},
\ee
and consider two regimes:

\medskip
\noindent{\bf Regime I}: $k$ fixed and $t\gg1$

This is the ``short-distance'' regime,
considering $k$ as the position of a fictitious random walker, as in section 1.
It is therefore analogous to the Porod regime
for phase-ordering systems~\cite{bray,glglau}.

Setting
\beq
f_k(t)\approx f_{k,\eq}\left(1+v_k\,t^{-\omega}\right),
\label{f1}
\eeq
equation~(\ref{df}) yields
\beq
v_k=v_0+A k,
\label{q1}
\eeq
where $v_0$ and $A$ are determined below.

\medskip
\noindent{\bf Regime II}: $k$ and $t$ simultaneously large (scaling regime)

In this regime, we look for a similarity solution
to~(\ref{df}) of the form
\beq
f_k(t)\approx f_{k,\eq}\,F(u),\qquad u=k\,t^{-1/2}.
\label{fsca}
\eeq
The structure of the master equations~(\ref{df})
indeed dictates that the scaling variable is the combination $kt^{-1/2}$.
Starting from a random initial condition, for a large but finite time~$t$,
and for $k$ much smaller than an ordering size of order $t^{1/2}$,
the system looks critical, i.e.,
the distribution $f_k(t)$ has essentially converged toward
the equilibrium distribution $f_{k,\eq}$.
This implies $F(0)=1$.
To the contrary, for $k\gg t^{1/2}$, the system still looks disordered,
i.e., the $f_k(t)$ fall off very fast.
Hence $F(\infty)=0$.
It will indeed be shown below that $f_k(t)\sim\exp(-k^2/(4t))$,
as the scaling function falls of very fast for $u\gg1$: $F(u)\sim\exp(-u^2/4)$.
Note the close analogy between the present situation and critical coarsening
for ferromagnetic spin systems~\cite{bray,glcrit},
where the scaling variable is $\abs{{\bf r}}t^{-1/z}$,
and where the role of $f_k(t)$ is played by the equal-time correlation
function $C(\abs{{\bf r}},t)$.

In order to determine the exponent $\omega$, we use
the sum rules~(\ref{sum1}) and~(\ref{sum2}), which yield respectively
\beqa
&&t^{-\omega}(v_0+A\rho_c)
=t^{-(\beta-1)/2}I_1,\qquad
I_1=\frac{1}{\zeta(\beta)}
\int_0^\infty u^{-\beta}\left(1-F(u)\right)\d u,\label{sumf1}\\
&&t^{-\omega}(v_0\rho_c+A\mu_c)
=t^{-(\beta-2)/2}I_2,\qquad
I_2=\frac{1}{\zeta(\beta)}
\int_0^\infty u^{1-\beta}\left(1-F(u)\right)\d u.\label{sumf2}
\eeqa
These equations are compatible only if the right-hand side
of equation~(\ref{sumf1}) is subleading.
In the case of a regular critical point ($\beta>3$), we thus obtain
\be
\omega=\frac{\beta-2}{2}>\frac{1}{2}
\ee
and
\be
A=\frac{I_2}{\mu_c-\rho_c^2},\qquad v_0=-\frac{\rho_c\,I_2}{\mu_c-\rho_c^2}.
\ee

Inserting the form~(\ref{fsca}) into~(\ref{df}),
and using the fact that $\omega>1/2$, leads to the differential equation
\beq
\D F(u)=0,
\label{fdif}
\eeq
where $\D$ is the linear differential operator
\beq
\D=-\frac{\d^2}{\d u^2}
+\left(-\frac{u}{2}+\frac{\beta}{u}\right)\frac{\d}{\d u}.
\label{ddef}
\eeq
The solution of~(\ref{fdif}) is
\beq
F(u)=\frac{2^{-\beta}}{\Gamma\!\left(\frac{\beta+1}{2}\right)}
\int_u^\infty y^\beta\,\e^{-y^2/4}\,\d y.
\label{fex}
\eeq

We present in Appendix A an alternative way of solving equation~(\ref{fdif}),
using the Mellin transformation.
In particular~(\ref{melumf}) yields the explicit expression
\be
I_2=\frac{M_{1-F}(\beta-2)}{\zeta(\beta)}=\frac{\pi^{1/2}\,2^{1-\beta}}
{(\beta-2)\,\Gamma\!\left(\frac{\beta+1}{2}\right)
\zeta(\beta)}.
\ee

As an illustration of the above,
let us determine how the variance of the population of box number 1,
$\Var N_1(t)=\mean{N_1(t)^2}-\mean{N_1(t)}^2=\mean{N_1(t)^2}-\rho_c^2$,
converges at long times to its equilibrium value $\mu_c-\rho_c^2$.
Only the scaling regime matters for the long-time behavior of this quantity,
and of all the quantities at criticality to be considered hereafter.
Using~(\ref{fsca}), we obtain
\be
\Var N_1(t)-(\mu_c-\rho_c^2)\approx-\Delta\,t^{-(\beta-3)/2},
\ee
with, by~(\ref{melumf}),
\be
\Delta=\frac{M_{1-F}(\beta-3)}{\zeta(\beta)}=\frac{2^{3-\beta}}
{(\beta-3)\,\Gamma\!\left(\frac{\beta+1}{2}\right)\zeta(\beta)}.
\ee

\subsection{In the condensed phase~($\rho>\rho_c$)}

We set
\be
\sigma(t)\approx 1+A\,t^{-1/2},
\ee
and consider the same two regimes as at criticality
(see~\cite{dgc} for more details).

\medskip
\noindent{\bf Regime~I}: $k$ fixed and $t\gg1$ (short-distance regime)

Equations~(\ref{f1}) and~(\ref{q1}) still hold,
but now with $\omega=1/2$, for any $\beta>2$.

\medskip
\noindent{\bf Regime~II}: $k$ and $t$ simultaneously large (scaling regime)

Looking for a similarity solution of equations~(\ref{df}), of the form
\beq
f_k(t)\approx\frac{F(u)}{t},\qquad u=k\,t^{-1/2},
\label{cofsca}
\eeq
we obtain for the scaling function $F(u)$ the linear differential
equation~\cite{dgc}
\beq
\frac{\d^2F}{\d u^2}
+\left(\frac{u}{2}-A+\frac{\beta}{u}\right)\frac{\d F}{\d u}
+\left(1-\frac{\beta}{u^2}\right)F=0.
\label{cofdif}
\eeq

Following~\cite{dgc}, we notice that the amplitude $A$
is determined by the fact that equation~(\ref{cofdif}) has an acceptable
solution $F(u)$, vanishing as $u\to0$ and $u\to\infty$.
The normalization of the solution $F(u)$ is determined by
the sum rule~(\ref{sum2}), which yields
\beq
\int_0^\infty u\,F(u)\,\d u=\rho-\rho_c.
\label{conorma}
\eeq
The parameter $v_0$ entering Regime~I is determined by the
sum rule~(\ref{sum1}), leading to
\be
v_0=-A\,\rho_c-\int_0^\infty F(u)\,\d u.
\ee

At variance with equation~(\ref{fdif}),
the differential equation~(\ref{cofdif}) cannot be solved in closed form.
It can be recast in Schwarzian form, without first derivative, by setting
\be
F(u)=u\,Y(u)\,H(u),
\ee
with
\beq
Y(u)=u^{-\beta/2-1}\,\e^{Au/2-u^2/8}.
\label{coeta}
\eeq
We thus obtain for $H(u)$ a differential equation of the form
\beq
\H\,H=0,
\label{cohdif}
\eeq
with
\beq
\H=-\frac{\d^2}{\d u^2}+W(u),
\label{cohdef}
\eeq
and where the potential $W(u)$ reads
\beq
W(u)=\frac{u^2}{16}-\frac{Au}{4}+\frac{\beta-3+A^2}{4}
-\frac{A\beta}{2u}+\frac{\beta(\beta+2)}{4u^2}.
\label{cow}
\eeq

Equation~(\ref{cohdif}) is a biconfluent Heun equation~\cite{heun}.
We will therefore refer to $\H$ as the Heun operator,
denoting its discrete eigenvalues by $E_n$
and the corresponding eigenfunctions bu $H_n(u)$.
Equation~(\ref{cohdif}) implies that the ground-state eigenvalue reads $E_0=0$,
while the associated eigenfunction $H_0(u)$
is simply related to the scaling function~$F(u)$:
\beq
F(u)=c\,u\,Y(u)\,H_0(u).
\label{coftoh}
\eeq
By~(\ref{conorma}) we have
\beq
c=\frad{\rho-\rho_c}{\int_0^{\infty^\vb}u^2\,Y(u)\,H_0(u)\,\d u}.
\label{coc}
\eeq

The spectrum of the operator $\H$,
and related quantities such as the scaling function $F(u)$,
can be further investigated in the limiting regimes of high and low
temperature.

\medskip
\noindent{\bf High temperature $(\beta\to0)$}

The analysis of this limiting situation will be helpful in section 7,
although it is of no direct physical relevance,
since the condensed phase only exists at low enough temperature $(\beta>2)$.

For $\beta=0$ and $A=0$, the potential~(\ref{cow}) becomes $W(u)=u^2/16-3/4$,
so that $\H$ is
the Hamiltonian of a harmonic oscillator on the half-line $u\ge0$,
up to a scale, with Dirichlet boundary condition at $u=0$.
Its spectrum is $E_n=n$ $(n=0,1,\dots)$,
and the (unnormalized) ground-state eigenfunction reads
\beq
H_0(u)=u\,\e^{-u^2/8},
\label{cohih}
\eeq
hence
\beq
F(u)=\frac{\rho-\rho_c}{2\,\pi^{1/2}}\,u\,\e^{-u^2/4}.
\eeq

Perturbation theory can then be used
to determine the small-$\beta$ behavior of the amplitude~$A$.
Expressing that the lowest eigenvalue reads $E_0=0$
with no correction, we obtain
\be
\int_0^\infty\left(-Au+\beta+\frac{2\beta}{u^2}\right)H_0^2(u)\,\d u+\cdots=0,
\ee
hence
\beq
A\approx\frac{\pi^{1/2}}{2}\beta.
\label{cohia}
\eeq

\medskip
\noindent{\bf Low temperature $(\beta\to\infty)$}

In this other limiting situation, the operator $\H$ simplifies as follows.
If we rescale $A$ and $u$ according to $A\sim u\sim\beta^{1/2}$,
all terms in the expression~(\ref{cow}) for the potential $W(u)$
scale as $\beta$.
In other words, $\beta$ plays the role of $1/\hbar$ in Quantum Mechanics,
and $\beta\to\infty$ is the semi-classical regime.
Expressing that the minimum of the potential is at zero yields
the first estimates $A\approx u\approx(2\beta)^{1/2}$.

A more refined analysis consists in expanding the potential $W(u)$
around its minimum.
Setting
\beq
A=(2\beta)^{1/2}-a,\qquad
u=(2\beta)^{1/2}+\beta^{1/8}v,
\label{coloset}
\eeq
the operator $\H$ becomes
\beq
\H\approx\frac{a^2-2}{4}
+\beta^{-1/4}\left(-\frac{\d^2}{\d v^2}+\frac{a}{2^{5/2}}v^2\right).
\label{cohred}
\eeq
The expression in the parentheses is again proportional
to the Hamiltonian of a harmonic oscillator.
Expressing that the lowest eigenvalue of the right-hand-side of~(\ref{cohred})
is $E_0=0$ determines $a=2^{1/2}-2^{-1/2}\beta^{-1/4}$, hence
\be
A\approx(2\beta)^{1/2}-2^{1/2}+2^{-1/2}\beta^{-1/4}.
\ee

The spectrum of $\H$ reads $E_n\approx\beta^{-1/4}\,n$ ($n=0,1,\dots$),
and the (unnormalized) ground-state eigenfunction reads
$H_0(u)\approx\e^{-v^2/4}$, i.e.,
\beq
H_0(u)\approx\exp\left(-\frac{(u-(2\beta)^{1/2})^2}{4\beta^{1/4}}\right).
\label{coloh}
\eeq
Both this expression and the function $Y(u)$ become singular
in the $\beta\to\infty$ limit,
so that a direct analysis of equation~(\ref{cofdif}) is needed
in order to derive
the behavior of the scaling function $F(u)$ at low temperature.
The above analysis suggests to look for a solution
depending on the scaling variable $y=(2\beta)^{-1/2}u$.
The term involving the second-order derivative $\d^2F/\d u^2$
is then negligible.
The simplified form of equation~(\ref{cofdif}), namely
\[
y(y-1)^2\,\frac{\d F}{\d y}+(2y^2-1)F=0,\qquad y=(2\beta)^{-1/2}u,
\]
admits the normalized solution
\[
F(u)=\frac{\rho-\rho_c}{2E_1(1)\,\beta}
\,\frac{y}{(1-y)^3}\,\exp\left(-\frac{1}{1-y}\right)\qquad(0<y<1),
\]
where $E_1$ is the first exponential integral $(E_1(1)=0.219383934)$.
The scaling function $F(u)$ is therefore nonzero in the $\beta\to\infty$ limit
only for $y<1$, i.e., $u<(2\beta)^{1/2}$.
The upper bound $y=1$, i.e., $u=(2\beta)^{1/2}$,
coincides with the point where the eigenfunction $H_0(u)$,
as given by equation~(\ref{coloh}), peaks at low temperature.

\section{Two-time observables: dynamical equations}

In this section we will successively establish dynamical equations for
the two-time correlation function $C(t,s)$,
for its derivative $\dpar C(t,s)/\dpar s$,
and for the response function $R(t,s)$.
(See~\cite{gl} for similar techniques.)

We consider the (connected) two-time correlation function $C(t,s)$
between the population of box number 1 at times
$s$~(waiting time) and $t$~(observation time), with $0\le s\le t$:
\be
C(t,s)=\mean{N_1(t)N_1(s)}-\mean{N_1(t)}\mean{N_1(s)}
=\mean{N_1(t)N_1(s)}-\rho^2.
\ee
This definition can be recast as~\cite{gl}
\be
C(t,s)=\sum_{k=1}^\infty k\,\g_k(t,s)-\rho^2,
\ee
with
\be
\g_k(t,s)=\sum_{j=1}^\infty j\,f_j(s)\,\P\{N_1(t)=k\mid N_1(s)=j\}.
\ee
The evolution of the $\g_k(t,s)$ with respect to $t$ is given, for $t\ge s$,
by a master equation similar to~(\ref{df}):
\beqa
&&\frad{\dpar\g_k(t,s)}{\dpar t}=\g_{k+1}(t,s)+\sigma(t)r_{k-1}\g_{k-1}(t,s)
-\left(1+\sigma(t)r_k\right)\g_k(t,s)\qquad(k\ge1),
\nonumber\\
&&\frad{\dpar\g_0(t,s)}{\dpar t}=\g_1(t,s)-\sigma(t)r_0\g_0(t,s).
\label{dg}
\eeqa
These equations preserve the sum rule $\sum_{k}\g_k(t,s)=\rho$.
At $t=s$, the initial conditions are $\g_k(s,s)=k\,f_k(s)$,
implying
$$
C(s,s)=\sum_{k=1}^\infty k^2\,f_k(s)-\rho^2,
$$
which is the variance of the population of box number 1
at time $s$, as it should.

In the discussion of the \fd theorem, we will
need expressions of the time derivative $\dpar C(t,s)/\dpar s$.
We have
\be
\frac{\dpar C(t,s)}{\dpar s}=\sum_{k=1}^\infty k\,\p_k(t,s),
\ee
with
\be
\p_k(t,s)=\frac{\dpar\g_k(t,s)}{\dpar s}.
\ee

The evolution of the $\p_k(t,s)$ with respect to $t$ is again given,
for $t\ge s$, by equations similar to~(\ref{df}):
\beqa
&&\frad{\dpar\p_k(t,s)}{\dpar t}=\p_{k+1}(t,s)+\sigma(t)r_{k-1}\p_{k-1}(t,s)
-\left(1+\sigma(t)r_k\right)\p_k(t,s)\qquad(k\ge1),
\nonumber\\
&&\frad{\dpar\p_0(t,s)}{\dpar t}=\p_1(t,s)-\sigma(t)r_0\p_0(t,s),
\label{dp}
\eeqa
with initial conditions at $t=s$
\bea
&&\p_k(s,s)=-f_{k+1}(s)+\sigma(s)r_{k-1}f_{k-1}(s)\qquad(k\ge1),
\nonumber\\
&&\p_0(s,s)=-f_1(s),
\eea
so that
\be
\left(\frac{\dpar C(t,s)}{\dpar s}\right)_{t=s}
=\sigma(s)\sum_{k=1}^\infty k\,r_k\,f_k(s)+2(1-f_0(s))-\rho.
\ee

The two-time response function $R(t,s)$
is a measure of the change in the mean population of box number 1 at time $t$,
induced by an infinitesimal modulation of the conjugate variable,
i.e., the local chemical potential acting on the same box,
at the earlier time $s$.

In the presence of an arbitrary local, time-dependent chemical potential
$\mu(t)$, the energy of box number 1 at time $t$ reads
\be
E\left(N_1(t)\right)=\ln(N_1(t)+1)-\mu(t)N_1(t).
\ee
The occupation probabilities of this box now depend on $\mu(t)$:
we denote them by $f_k^\mu(t)$.
In the thermodynamic limit, i.e., to leading order as $M\to\infty$,
the occupation probabilities of all
the other boxes $(i=2,\dots,M)$ are still given by the $f_k(t)$.

The response function reads
\be
R(t,s)=\left(\frac{\delta\mean{N_1(t)}}{\delta\mu(s)}\right)_{\mu=0}
=\sum_{k=1}^\infty k\,\h_k(t,s),
\ee
with
\be
\h_k(t,s)=\left(\frac{\delta f^\mu_k(t)}{\delta\mu(s)}\right)_{\mu=0}.
\ee

The modified occupation probabilities $f_k^\mu(t)$ obey the dynamical equations
\beqa
&&\frad{\d f^\mu_k(t)}{\d t}
=f^\mu_{k+1}(t)+\sigma(t)\,\e^{\beta\mu(t)}r_{k-1}f^\mu_{k-1}(t)
-\left(1+\sigma(t)\,\e^{\beta\mu(t)}r_k\right)f^\mu_k(t)\qquad(k\ge1),
\nonumber\\
&&\frad{\d f^\mu_0(t)}{\d t}
=f^\mu_1(t)-\sigma(t)\,\e^{\beta\mu(t)}r_0f^\mu_0(t).
\label{dfa}
\eeqa
The initial values $f^\mu_k(0)=f_k(0)$ and the parameter $\sigma(t)$
are unchanged.
The dynamical equations~(\ref{dfa}) preserve the sum rule
$\sum_k f^\mu_k(t)=1$.

Equations~(\ref{dfa}) imply that the $\h_k(t,s)$ obey, for $t>s$,
\beqa
&&\frad{\dpar\h_k(t,s)}{\dpar t}=\h_{k+1}(t,s)+\sigma(t)r_{k-1}\h_{k-1}(t,s)
-\left(1+\sigma(t)r_k\right)\h_k(t,s)\qquad(k\ge1),
\nonumber\\
&&\frad{\dpar\h_0(t,s)}{\dpar t}=\h_1(t,s)-\sigma(t)r_0\h_0(t,s),
\label{dh}
\eeqa
with initial conditions at $t=s$
\bea
&&\h_k(s,s)
=\beta\sigma(s)\left(r_{k-1}f_{k-1}(s)-r_kf_k(s)\right)\qquad(k\ge1),
\nonumber\\
&&\h_0(s,s)=-\beta\sigma(s)r_0f_0(s),
\eea
so that
\be
R(s,s)=\beta\left(1-f_0(s)\right).
\ee

The behavior of the two-time observables will now be successively investigated
in the next three sections, first at criticality~($\rho=\rho_c$),
both at equilibrium and in the nonequilibrium regime,
and then in the condensed phase~($\rho>\rho_c$).

\section{Equilibrium critical dynamics}

When the waiting time $s$ becomes very large,
keeping the difference $\tau=t-s\ge0$ fixed,
two-time quantities reach their equilibrium values,
which only depend on $\tau$, both in the fluid phase ($\rho<\rho_c$)
and along the critical line ($\rho=\rho_c$).
This section is devoted to the latter case.

The equilibrium correlation function $C_\eq(\tau)$ reads
\be
C_\eq(\tau)=\sum_{k=1}^\infty k\,\g_{k,\eq}(\tau)-\rho_c^2,
\ee
where the equilibrium values $\g_{k,\eq}(\tau)$ of the $\g_k(t,s)$ obey
\beqa
&&\frad{\d\g_{k,\eq}(\tau)}{\d t}
=\g_{k+1,\eq}(\tau)+r_{k-1}\g_{k-1,\eq}(\tau)
-\left(1+r_k\right)\g_{k,\eq}(\tau)
\qquad(k\ge1),
\nonumber\\
&&\frad{\d\g_{0,\eq}(\tau)}{\d t}
=\g_{1,\eq}(\tau)-r_0\g_{0,\eq}(\tau),
\label{dgeq}
\eeqa
with initial conditions $\g_{k,\eq}(0)=k\,f_{k,\eq}$.
This implies consistently that $C_\eq(0)=\mu_c-\rho_c^2$
is the equilibrium variance of the population of a generic box.

As the time difference $\tau$ gets large,
we have $\g_{k,\eq}(\tau)\to\rho_c\,f_{k,\eq}$, so that $C_\eq(\tau)\to0$.
The decay of $C_\eq(\tau)$ for large $\tau$ can be investigated along the
lines of section~3.1.
The two regimes I and II are again to be considered separately,
although results concerning the former will not be needed explicitly.
In the scaling regime, we look for a similarity solution
to equations~(\ref{dgeq}) of the form
\beq
\g_{k,\eq}(\tau)\approx
f_{k,\eq}\,\tau^{1/2}\,G_\eq(u),\qquad u=k\,\tau^{-1/2},
\label{geqsca}
\eeq
for which equations~(\ref{dgeq}) yield the differential equation
\beq
\left(\D+\frac12\right)G_\eq(u)=0,
\label{geqdif}
\eeq
with boundary conditions $G_\eq(0)=0$ and $G_\eq(u)\approx u$ as $u\to\infty$.
The differential operator $\D$ was defined in equation~(\ref{ddef}).

Equation~(\ref{geqdif}) can be solved by the method of Appendix A.
With the definition~(\ref{mdef}), we have the functional equation
\be
\frac{M_{G_\eq}(z+2)}{M_{G_\eq}(z)}=\frac{z-1}{2(z+2)(\beta-1-z)},
\ee
whose suitably normalized solution reads
\be
M_{G_\eq}(z)=\frac
{\pi^{1/2}\,\Gamma\!\left(\frac{z-1}{2}\right)\,
\Gamma\!\left(\frac{\beta+1-z}{2}\right)}
{2^{z+1}\,\Gamma\!\left(\frac{z+2}{2}\right)\,
\Gamma\!\left(\frac{\beta}{2}\right)}
\qquad(1<\re z<\beta+1).
\ee

The decay of the equilibrium correlation function at large $\tau$
is again dominated by the scaling regime, and the scaling form~(\ref{geqsca})
yields
\beq
C_\eq(\tau)\approx A_\eq\,\tau^{-(\beta-3)/2},
\label{ceq}
\eeq
with
\be
A_\eq=\frac{M_{G_\eq}(\beta-2)}{\zeta(\beta)}
=\frac{\pi\,\Gamma\!\left(\frac{\beta-3}{2}\right)}
{2^\beta\,\Gamma\!\left(\frac{\beta}{2}\right)^2\zeta(\beta)}.
\ee

As expected, the \fd theorem holds at equilibrium.
Indeed the equilibrium values $\p_{k,\eq}(\tau)$ and $\h_{k,\eq}(\tau)$
of $\p_k(t,s)$ and $\h_k(t,s)$
obey the same equations, identical to equations~(\ref{dgeq}),
with initial conditions:
\bea
&&\h_{k,\eq}(0)=\beta\p_{k,\eq}(0)=
\beta\left(f_{k,\eq}-f_{k+1,\eq}\right)\qquad(k\ge1),
\nonumber\\
&&\h_{0,\eq}(0)=\beta\p_{0,\eq}(0)=-\beta f_{1,\eq},
\eea
hence the identities
\beq
\h_{k,\eq}(\tau)=\beta\p_{k,\eq}(\tau)
\eeq
and
\be
R_\eq(\tau)=-\beta\,\frac{\d C_\eq(\tau)}{\d\tau}.
\ee
The last formula is the \fd theorem in its usual differential form.

\section{Nonequilibrium critical dynamics}

We now turn to the nonequilibrium critical behavior
of the two-time correlation function $C(t,s)$,
response function $R(t,s)$, and \fd ratio~\cite{ck,aging}
\beq
X(t,s)=\frac{R(t,s)}{\beta\,\frad{\dpar C(t,s)}{\dpar s}},
\label{xdef}
\eeq
in the scaling regime where both time variables $s$ and $t$
are large and comparable.
Hereafter $x$ will denote the dimensionless time ratio
\be
x=\frac{t}{s}\ge1.
\ee

We first analyze the scaling behavior of the correlation function $C(t,s)$.
Looking for a two-variable scaling solution to equations~(\ref{dg})
of the form
\beq
\g_k(t,s)\approx f_{k,\eq}\,t^{1/2}\,G(u,x),\qquad u=k\,t^{-1/2},\qquad x=t/s,
\label{gsca}
\eeq
we obtain the partial differential equation
\beq
\left(x\frac{\dpar}{\dpar x}+\D+\frac12\right)G(u,x)=0
\label{gdif}
\eeq
for the scaling function $G(u,x)$, with initial condition
\beq
G(u,1)=u F(u),
\label{initg}
\eeq
and boundary condition $G(\infty,x)=0$ for all $x\ge1$.
The function $F(u)$ is known from~(\ref{fex}),
and the operator $\D$ is given by~(\ref{ddef}).

As shown in Appendix B, equation~(\ref{gdif})
can be solved explicitly by the method of separation of variables.
Equation~(\ref{gsca}) implies the scaling law
\beq
C(t,s)\approx s^{-(\beta-3)/2}\,\Phi(x),
\label{csca}
\eeq
where
\be
\Phi(x)=\frac{x^{-(\beta-3)/2}}{\zeta(\beta)}\int_0^\infty
u^{1-\beta}\,G(u,x)\,\d u.
\ee
Using~(\ref{gexpand}), we obtain
\beq
\Phi(x)=x^{-\beta/2}\sum_{n=0}^\infty A_n\,x^{-n},
\label{phiexp}
\eeq
with
\be
A_n=\frac{a_n\,M_{G_n}(\beta-2)}{\zeta(\beta)}.
\ee
More explicitly, equations~(\ref{melgn}) and~(\ref{an}) imply that
the leading coefficient of the expansion~(\ref{phiexp}) reads
\beq
A_0=\frac{\pi^{1/2}\,2^{3-\beta}\Gamma\!\left(\frac{\beta+4}{2}\right)}
{3(\beta+1)\,\Gamma\!\left(\frac{\beta+1}{2}\right)^2\zeta(\beta)},
\eeq
while the coefficient ratios are rational functions of $\beta$:
\beq
\frac{A_n}{A_0}
=\frac{3}{2^n}\prod_{j=0}^{n-1}(2j+\beta)\,\cdot\,
\sum_{k=0}^n\frac{(-)^k}{(2k+3)k!(n-k)!}\prod_{\ell=0}^{k-1}
\frac{2\ell+\beta+4}{2\ell+\beta+3},
\eeq
i.e.,
\be
\frac{A_1}{A_0}=\frac{\beta(2\beta+3)}{10(\beta+3)},\qquad
\frac{A_2}{A_0}=\frac{\beta(\beta+2)(8\beta^2+52\beta+45)}
{280(\beta+3)(\beta+5)},\qquad\hbox{etc.}
\ee

The expansion~(\ref{phiexp})
is convergent over the whole physical domain $(x>1)$.
For $x\to1$, i.e., $\tau=t-s\ll s$, the equilibrium result~(\ref{ceq})
is recovered as
\beq
\Phi(x)\approx A_\eq(x-1)^{-(\beta-3)/2}.
\label{phieq}
\eeq
These properties can be checked by noticing that the expression~(\ref{lag})
for the Laguerre polynomials simplifies to
\be
\L_n(u^2/4)\approx\left(\frac{2\,n^{1/2}}{u}\right)^{(\beta+1)/2}
\J\left(u\,n^{1/2}\right),
\ee
where $J$ is the Bessel function,
in the scaling regime where the order $n$ is large and $u$ is small.
The subsequent integrals can be estimated for $n$ large, yielding
\be
A_n\approx
\frac{\pi}{2^\beta\,\Gamma\!\left(\frac{\beta}{2}\right)^2\zeta(\beta)}
\,n^{(\beta-5)/2}.
\ee
This asymptotic expression ensures the convergence
of the series~(\ref{phiexp}) for all $x>1$,
and establishes~(\ref{phieq}), including the prefactor.

We now turn to the derivative $\dpar C(t,s)/\dpar s$
of the correlation function.
Looking for a two-variable scaling solution to equations~(\ref{dp})
in the scaling regime, of the form
\beq
\p_k(t,s)\approx f_{k,\eq}\,t^{-1/2}\,G\un(u,x),
\label{g1sca}
\eeq
we get the partial differential equation
\beq
\left(x\frac{\dpar}{\dpar x}+\D-\frac12\right)G\un(u,x)=0,
\label{g1dif}
\eeq
with initial condition
\beq
G\un(u,1)=\beta\,\frac{F(u)}{u}-2F'(u).
\label{initg1}
\eeq

Equation~(\ref{g1dif}) can again be solved by the method of Appendix B.
We thus obtain
\beq
\frac{\dpar C(t,s)}{\dpar s}\approx s^{-(\beta-1)/2}\,\Phi\un(x),
\label{dcsca}
\eeq
where
\beq
\Phi\un(x)=\frac{x^{-(\beta-1)/2}}{\zeta(\beta)}\int_0^\infty
u^{1-\beta}\,G\un(u,x)\,\d u.
\label{intphi1}
\eeq
This expression will not be made more explicit,
as equation~(\ref{csca}) yields more directly
\beq
\Phi\un(x)=\frac{3-\beta}{2}\Phi(x)-x\,\frac{\d\Phi}{\d x},
\label{phiphi1}
\eeq
i.e.,
\beq
\Phi\un(x)=x^{-\beta/2}\sum_{n=0}^\infty A\un_n\,x^{-n},
\label{phi1exp}
\eeq
with
\beq
A\un_n=\left(n+\frac32\right)A_n.
\eeq

The scaling behavior of the response function $R(t,s)$
can be determined by the same approach.
Looking for a two-variable scaling solution to equations~(\ref{dh}) in the
scaling
regime, of the form
\be
\h_k(t,s)\approx f_{k,\eq}\,t^{-1/2}\,G\de(u,x),
\ee
equations~(\ref{dh}) yield the partial differential equation
\beq
\left(x\frac{\dpar}{\dpar x}+\D-\frac12\right)G\de(u,x)=0,
\label{g2dif}
\eeq
with initial condition
\beq
G\de(u,1)=\beta^2\frac{F(u)}{u}-\beta F'(u).
\label{initg2}
\eeq

We thus obtain
\beq
R(t,s)\approx s^{-(\beta-1)/2}\,\Phi\de(x),
\label{rsca}
\eeq
where
\beq
\Phi\de(x)=\frac{x^{-(\beta-1)/2}}{\zeta(\beta)}\int_0^\infty
u^{1-\beta}\,G\de(u,x)\,\d u,
\label{intphi2}
\eeq
i.e., explicitly
\beq
\Phi\de(x)=x^{-\beta/2}\sum_{n=0}^\infty A\de_n\,x^{-n}.
\label{phi2exp}
\eeq
The leading coefficient of this expansion reads
\be
A\de_0=\frac{\beta\,\pi^{1/2}\,2^{1-\beta}
\Gamma\!\left(\frac{\beta+2}{2}\right)}
{\Gamma\!\left(\frac{\beta+1}{2}\right)^2\zeta(\beta)}
=\frac{3\beta(\beta+1)}{2(\beta+2)}\,A_0,
\ee
while the coefficient ratios are again rational functions of $\beta$:
\be
\frac{A\de_n}{A\de_0}
=\frac{1}{(\beta+1)2^n}\prod_{j=0}^{n-1}(2j+\beta)\,\cdot\,
\sum_{k=0}^n\frac{(-)^k(2k+\beta+1)}{(2k+1)k!(n-k)!}\prod_{\ell=0}^{k-1}
\frac{2\ell+\beta+2}{2\ell+\beta+3},
\ee
i.e.,
\be
\frac{A\de_1}{A\de_0}=\frac{\beta(2\beta+1)}{6(\beta+1)},\qquad
\frac{A\de_2}{A\de_0}=\frac{\beta(\beta+2)(8\beta^2+28\beta+9)}
{120(\beta+1)(\beta+3)},\qquad\hbox{etc.}
\ee

The scaling results~(\ref{dcsca}) and~(\ref{rsca})
imply that the \fd ratio $X(t,s)$, defined in equation~(\ref{xdef}),
only depends on the time ratio $x$ in the nonequilibrium scaling regime.
We have indeed
\beq
X(t,s)\approx\X(x)=\frac{\Phi\de(x)}{\beta\Phi\un(x)}.
\label{xsca}
\eeq
The scaling function $\X(x)$ turns out to be universal,
whereas $\Phi(x)$, $\Phi\un(x)$, and $\Phi\de(x)$,
which respectively enter equations~(\ref{csca}),~(\ref{dcsca}),~(\ref{rsca}),
are only universal up to a scale fixing.
(A definition of universal quantities
in the present context will be recalled in the Discussion.)

Equations~(\ref{phi1exp}) and~(\ref{phi2exp}) yield the expression
\beq
\X(x)
=\frad{\sum_{n=0}^\infty A\de_n\,x^{-n}}{\beta\sum_{n=0}^\infty A\un_n\,x^{-n}}
=\sum_{n=0}^\infty\xi_n\,x^{-n},
\label{xexp}
\eeq
which is again convergent for $x>1$.
The limit \fd ratio, $X_\infty=\X(\infty)=\xi_0$, takes the simple value
\beq
X_\infty=\frac{\beta+1}{\beta+2}.
\label{xinf}
\eeq
Equation~(\ref{xexp}) also yields
\be
\xi_1=\frac{\beta^2}{3(\beta+2)(\beta+3)},\qquad
\xi_2=\frac{\beta^2(-2\beta^3+2\beta^2+78\beta+117)}
{45(\beta+2)(\beta+3)^2(\beta+5)},\qquad\hbox{etc.}
\ee

The behavior of the \fd ratio $\X(x)$
close to equilibrium (i.e., for $x\to1$) is also of interest.
Equations~(\ref{intphi1}) and~(\ref{intphi2}) for $x=1$,
together with~(\ref{initg1}) and~(\ref{initg2}), imply
\be
\lim_{x\to1}\left(\beta\Phi\un(x)-\Phi\de(x)\right)
=-\frac{\beta}{\zeta(\beta)}\,M_{F'}(\beta-2)
=\frac{2^{1-\beta}\,\beta}{\Gamma\!\left(\frac{\beta+1}{2}\right)\zeta(\beta)}.
\ee
This expression, together with equations~(\ref{phieq}) and~(\ref{phiphi1}),
yields
\be
\X(x)\approx1-\frac{2\,\Gamma\!\left(\frac{\beta}{2}\right)^2}
{\pi\,\Gamma\!\left(\frac{\beta+1}{2}\right)
\Gamma\!\left(\frac{\beta-1}{2}\right)}\,(x-1)^{(\beta-1)/2},
\ee
confirming that the \fd theorem is restored for $x\to1$,
and explicitly giving the leading violation of that theorem
in the scaling regime, which is proportional to $(\tau/s)^{(\beta-1)/2}$.

\section{Nonequilibrium dynamics in the condensed phase}

We finally turn to the long-time behavior of the two-time quantities
$C(t,s)$, $R(t,s)$, and $X(t,s)$ in the condensed phase~($\rho>\rho_c$).

In Regime~II, we look for two-variable scaling solutions
to equations~(\ref{dg}),~(\ref{dp}), and~(\ref{dh}),
inspired by~(\ref{cofsca}), of the form
\beq
\g_k(t,s)\approx\frac{k}{t}\,G(u,x),\quad
\p_k(t,s)\approx\frac{k}{t^2}\,G\un(u,x),\quad
\h_k(t,s)\approx\frac{k}{t^2}\,G\de(u,x),
\label{cosca}
\eeq
where $u=k\,t^{-1/2}$ and $x=t/s$.

The three scaling functions obey partial differential equations
of the form~(\ref{gdif}),~(\ref{g1dif}), or~(\ref{g2dif}),
where the differential operator $\D$ is replaced by
\beq
\D=-\frac{\d^2}{\d u^2}
+\left(-\frac{u}{2}+A-\frac{\beta+2}{u}\right)\frac{\d}{\d u}
-\frac{3}{2}+\frac{A}{u}.
\eeq
This differential operator is related to the Heun operator $\H$~(\ref{cohdef})
and to the function $Y(u)$ defined in equation~(\ref{coeta}) by the conjugation
\beq
Y(u)\,\H=\D\,Y(u).
\eeq
Hence $\D$ and $\H$ share the same eigenvalues $E_n$ ($n=0,1,\dots$),
which are not known explicitly, as already mentioned, except $E_0=0$.

The two-time scaling functions defined in~(\ref{cosca})
are then determined in analogy with the critical case.
We thus obtain
\beq
G(u,x)=Y(u)\sum_{n=0}^\infty a_n x^{-E_n-1/2}\,H_n(u),\qquad
G\unde(u,x)=Y(u)\sum_{n=0}^\infty a\unde_n x^{-E_n+1/2}\,H_n(u),
\label{coexpan}
\eeq
where the coefficients $a_n$, $a\unde_n$ are determined
by the initial conditions
\beq
G(u,1)=F(u),\quad
G\un(u,1)=\left(\frac{A}{u}-\frac{\beta}{u^2}\right)F(u)
-\frac{2}{u}\,F'(u),\quad
G\de(u,1)=-\frac{\beta}{u}\,F'(u).
\label{cotwoi}
\eeq

Using the expansions~(\ref{coexpan}),
as well as the linear dependence of $F(u)$ in $(\rho-\rho_c)$
(see equations~(\ref{coftoh}),~(\ref{coc})),
we obtain the following scaling forms for the
correlation and response functions:
\beq
\matrix{
\ds{C(t,s)\approx s^{1/2}\,(\rho-\rho_c)\,\Phi(x),}\hfill&
\ds{\Phi(x)=\sum_{n=0}^\infty A_n\,x^{-E_n},}\hfill\cr
\ds{\frac{\dpar C(t,s)}{\dpar s}
\approx s^{-1/2}\,(\rho-\rho_c)\,\Phi\un(x),}\qquad\hfill&
\ds{\Phi\un(x)=\sum_{n=0}^\infty A\un_n\,x^{-E_n},}\hfill\cr
\ds{R(t,s)\approx s^{-1/2}\,(\rho-\rho_c)\,\Phi\de(x),}\hfill&
\ds{\Phi\de(x)=\sum_{n=0}^\infty A\de_n\,x^{-E_n}.}\hfill\cr
}
\eeq
The scaling functions $\Phi(x)$, $\Phi\unde(x)$ only depend on $\beta$.
They take finite limit values, both at $x=0$ and at $x=\infty$.
The \fd ratio
\be
X(t,s)\approx\X(x)=\frac{\Phi\de(x)}{\beta\Phi\un(x)}
\ee
again only depends on the time ratio $x$
in the nonequilibrium scaling regime.
The whole scaling function $\X(x)$ is universal,
just as in the critical case.
In particular, the limit \fd ratio $X_\infty=\X(\infty)=A\de_0/A\un_0$
in the condensed phase is universal, and only depends on $\beta$.
Using equations~(\ref{coftoh}) and~(\ref{cotwoi}),
we can derive the following expression for $X_\infty$
in terms of the ground-state eigenfunction $H_0(u)$:
\beq
X_\infty=\frad
{\int_0^\infty\left(\frac{u}{2}-A+\frac{\beta}{u}\right) H_0^2(u)\,\d u}
{\int_0^{\infty^\vb}u\,H_0^2(u)\,\d u}.
\label{coxinf}
\eeq

The above expressions cannot be made more explicit in general.
More quantitative results can be derived
at high temperature and at low temperature, using results of section~3.2.

Let us consider the limit \fd ratio $X_\infty$, as given by
equation~(\ref{coxinf}).
In the high-temperature case, by~(\ref{cohih}) and~(\ref{cohia}), we obtain
\beq
X_\infty=\frac{1}{2}-(\pi-2)\beta+\cdots\qquad(\beta\to 0).
\label{coxhi}
\eeq
In the low-temperature case, using~(\ref{coloset}) and~(\ref{coloh}),
we obtain
\beq
X_\infty=\beta^{-1/2}-\frac{\beta^{-3/4}}{4}+\cdots\qquad(\beta\to\infty).
\label{coxlo}
\eeq

\begin{figure}[htb]
\begin{center}
\includegraphics[angle=90,width=.8\linewidth]{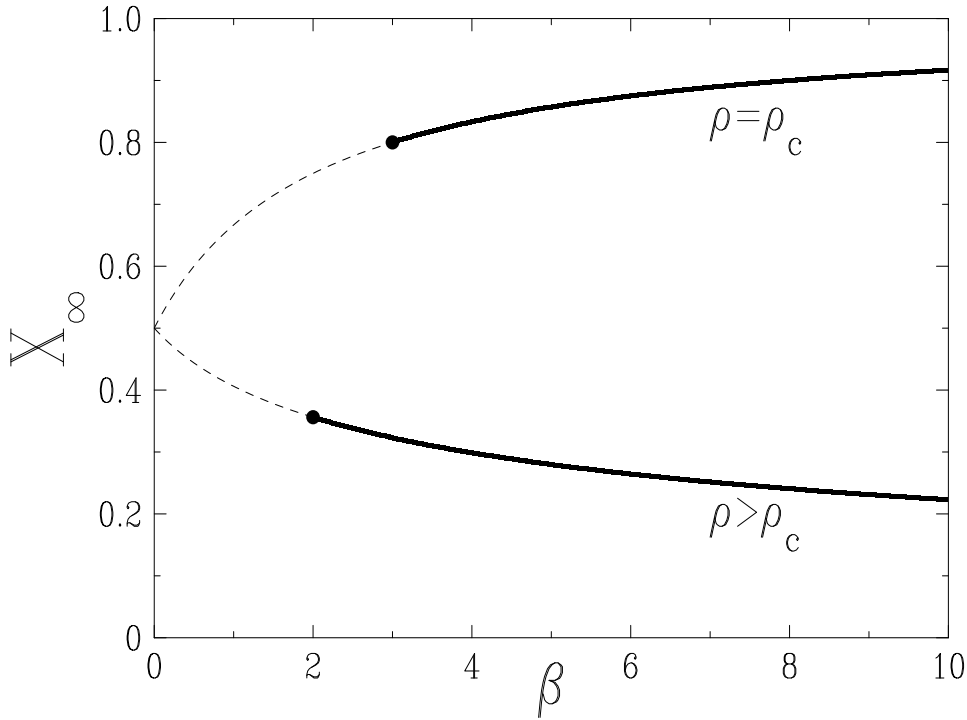}
\caption{\small
Plot of the limit \fd ratio $X_\infty$ against inverse temperature $\beta$.
Upper curve: critical point ($\beta>3$, $\rho=\rho_c$)
(see~(\ref{xinf})).
Lower curve: condensed phase ($\beta>2$, $\rho>\rho_c$)
(see~(\ref{coxinf})).
Thin dashed lines: continuation of the results to high temperature.}
\label{fx}
\end{center}
\end{figure}

It is instructive to compare the limit \fd ratios
corresponding to the critical point and to the condensed phase.
The value of $X_\infty$ at criticality is given by the analytical
expression~(\ref{xinf}).
The value of $X_\infty$ in the condensed phase is obtained by
a numerical evaluation of~(\ref{coxinf}):
the ground-state eigenfunction $H_0(u)$ is obtained by numerically
solving the differential equation~(\ref{cofdif})
or~(\ref{cohdif}), and then used to evaluate
the integrals entering the result~(\ref{coxinf}).
The values thus obtained smoothly interpolate
between the limiting laws~(\ref{coxhi}) and~(\ref{coxlo}).
In Figure~\ref{fx},
thick full lines show the physical values of the \fd ratios,
while thin dashed lines show their continuation to high temperatures.
Both \fd ratios start from $1/2$ at infinite temperature,
and converge to the limit values $X_\infty=0$ and $X_\infty=1$
at zero temperature, respectively in the condensed phase and at criticality
(see Discussion).

\section{Discussion}

At the onset of this paper we gave a comparative presentation
of two classes of dynamical urn models,
the Ehrenfest class and the Monkey class.
All these models have simple static properties,
because their Hamiltonian is a sum of contributions of independent boxes.
They possess however interesting nonequilibrium dynamical properties,
even in the mean-field geometry.
The backgammon model~\cite{ritort,fr,gbm,gl}
is a prototypical example of the Ehrenfest class, while
model B of Reference~\cite{gbm}, and the zeta urn model~\cite{bia,dgc}
which is the subject of the present work, are examples of the other class.

Let us come back to the role of statistics in the definition and
properties of such models.
The essential difference between the Ehrenfest class and the Monkey class
indeed resides in matters related to a priori statistics.
Statistics enters the dynamical definition of models:
the proposed moves for the Ehrenfest class (respectively, the Monkey class)
are chosen according to ball-box statistics (respectively, box-box statistics).
Consistently, statistics also enters their equilibrium definition:
the partition functions~(\ref{z2n}),~(\ref{zinfty})
of the original Ehrenfest model, and of its generalization to $M$ urns,
involve a factorial of the total number of particles in their denominators.
Inverse factorials $1/N_i!$,
taking into account equivalent labelings of particles within each box,
are also involved in the evaluation of the partition function~(\ref{jstar}).

In statistical mechanics with Maxwell-Boltzmann statistics,
the presence of inverse factorials has its origin in
the indiscernibility of identical classical objects.
These factorials are absent for the Monkey class,
in which the populations $N_i$ are involved in flat sums,
such as~(\ref{Zbb}), just as occupation numbers are in quantum-mechanical
statistical mechanics with Bose-Einstein statistics.
There is of course nothing quantum-mechanical in the urn models considered
here.
It can be said for short, following Reference~\cite{kim},
that the equilibrium statistics of the Ehrenfest class is Maxwell-Boltzmann,
while that of the Monkey class is Bose-Einstein.
Table~1 illustrates this discussion.

\begin{table}[htb]
\begin{center}
\begin{tabular}{|c|c|c|}
\hline
Class&{\bf Ehrenfest}&{\bf Monkey}\\
\hline
Dynamical rule&ball-box&box-box\\
\hline
Equilibrium weight of box
$i$&$\frad{p_{N_i}^\vb}{N_{i_{\vb_\vb}}!}$&$p_{N_i}$\\
\hline
Statistical-mechanical analogue&classical&quantum-mechanical\\
&(Maxwell-Boltzmann)&(Bose-Einstein)\\
\hline
\end{tabular}
\caption{\small
Comparison of a priori statistics for the Ehrenfest class and the Monkey class
of dynamical urn models: dynamical and equilibrium aspects.}
\end{center}
\end{table}

The main focus of this work concerns the nonequilibrium properties
of the zeta urn model,
with emphasis on the aging behavior of the two-time correlation
and response functions of the fluctuating population of a given box,
and of the corresponding \fd ratio.
We considered successively the critical line ($\rho=\rho_c$) and the condensed
phase ($\rho>\rho_c$).
We summarize these two cases below.

The critical line in the temperature-density plane
corresponds to a line of fixed points,
parametrized by inverse temperature $\beta$~\cite{bia}.
In other words, critical exponents, both static and dynamical,
and more generally universal quantities, depend continuously on temperature.
In the present context, a universal quantity ought to be independent:
\begin{description}
\item
-- of the initial state (provided it is homogeneous and disordered,
i.e., the $f_k(0)$ decay rapidly),
\item
-- of the specific form of the energy of each box
(provided it diverges logarithmically, as $E(N_i)\approx\ln N_i$,
for large occupation numbers),
\item
-- and of details of the dynamics (such as Metropolis versus heat-bath).
\end{description}

Our results~(\ref{csca}) and~(\ref{rsca})
for the two-time correlation and response function
have the expected product form~(see~\cite{glcrit} and references therein),
involving: a common non-universal prefactor,
a negative power of the waiting time, related to the anomalous dimension
of the observable, and a universal scaling function of the time ratio,
or temporal aspect ratio, $x=t/s$.
Accordingly, the \fd ratio~(\ref{xsca}), $X(t,s)\approx\X(x)$,
only depends on $x$ in the nonequilibrium scaling regime.
The limit value $X_\infty=\X(\infty)$ has been recently
emphasized~\cite{glglau,glcrit} to be a new universal quantity,
characteristic of nonequilibrium critical dynamics.
For the zeta urn model, the result~(\ref{xinf})
applies to the regular part of the critical line~($\beta>3$),
so that $4/5<X_\infty<1$, as shown by the upper curve in Figure~\ref{fx}.
This range is unusual for a critical system.
Indeed, statistical-mechanical models such as ferromagnets
are observed to have $0<X_\infty\le1/2$ at their critical point.
The upper bound $X_\infty=1/2$, corresponding to the mean-field
situation~\cite{glcrit},
is also observed in a range of simpler models~\cite{ck,aging,glglau}.
It is worth noticing that the backgammon model
also has a high \fd ratio at its zero-temperature critical point,
namely $X(t,s)\approx 1-C/(\ln s)^2$ for $s\ll t$,
where the amplitude $C$ depends both on the observable
and on the dynamical rule~\cite{fr,gl}.

The present analysis of condensation dynamics
for $\rho>\rho_c$ extends and completes that begun in~\cite{dgc}.
We have investigated various quantities related to the occupation probabilities
in the scaling regime, describing the growing condensate.
An approximate analysis of the Heun operator~(\ref{cohdef})
has allowed us to obtain asymptotic expressions
for various quantities at high and low temperature.
The \fd ratio admits a non-trivial limit value $X_\infty$
throughout the condensed phase, shown by the lower curve in Figure~\ref{fx}.
Expression~(\ref{coxlo}) shows that $X_\infty\approx\beta^{-1/2}$
slowly goes to zero at low temperature.
This behavior is very different from that of conventional models.
It is indeed currently accepted~\cite{xzero} that coarsening systems,
such as ferromagnets quenched from a high-temperature initial state,
have identically $X_\infty=0$ throughout their low-temperature phase,
i.e., for any temperature below $T_c$.
These unusual features of the zeta urn model are less of a surprise
if one remembers that the condensation dynamics of the present model
is basically different from a domain-growth or coarsening dynamics.
In the latter case, phase separation takes place in a statistically homogeneous way, at least for an infinite system.
To the contrary, in the present situation,
condensation takes place in a very inhomogeneous fashion.
The form of equation~(\ref{cofsca}) indeed demonstrates that
the condensate of particles is shared
by an ever decreasing fraction of boxes, scaling as $t^{-1/2}$,
each of them having a population growing as $t^{1/2}$
(until size effects eventually become important for $t\sim M^2$).

\newpage
\appendix
\section{Solving equation~(\ref{fdif}) by Mellin transformation}

The Mellin transformation provides an efficient alternative way of solving
the differential equation~(\ref{fdif}).

We define the Mellin transform $M_f(z)$ of a function $f(u)$ by
\beq
M_f(z)=\int_0^\infty u^{-z-1}\,f(u)\,\d u,\qquad
f(u)=\int\frac{\d z}{2\pi\i}\,u^z\,M_f(z).
\label{mdef}
\eeq

Equation~(\ref{fdif}), together with the boundary condition $F(0)=1$,
is equivalent to the following functional equation
\be
\frac{M_{1-F}(z+2)}{M_{1-F}(z)}=\frac{z}{2(z+2)(\beta-1-z)}
\ee
for the Mellin transform of $1-F(u)$.
The solution of this functional equation,
with boundary condition $F(\infty)=0$, reads
\beq
M_{1-F}(z)=\frac{\Gamma\!\left(\frac{\beta+1-z}{2}\right)}
{z\,2^z\,\Gamma\!\left(\frac{\beta+1}{2}\right)}\qquad(0<\re z<\beta+1).
\label{melumf}
\eeq
Similarly, for $\re z<0$, the Mellin transform of $F(u)$,
\beq
M_F(z)=-\frac{\Gamma\!\left(\frac{\beta+1-z}{2}\right)}
{z\,2^z\,\Gamma\!\left(\frac{\beta+1}{2}\right)}\qquad(\re z<0),
\label{melf}
\eeq
is the formal opposite of equation~(\ref{melumf}).

We also give for further reference the expression of the Mellin transform
of the derivative $F'(u)$:
\beq
M_{F'}(z)=-\underbrace{(z+1)M_{1-F}(z+1)}_{-1<\re z<\beta}
=\underbrace{(z+1)M_F(z+1)}_{\re z<-1}
=-\frac{\Gamma\!\left(\frac{\beta-z}{2}\right)}
{2^{z+1}\,\Gamma\!\left(\frac{\beta+1}{2}\right)}
\qquad(\re z<\beta).
\label{melfp}
\eeq

\section{Solving equation~(\ref{gdif}) by separation of variables
and spectral superposition}

Partial differential equations such as~(\ref{gdif}),
with given initial and boundary conditions,
can be solved explicitly by the method of separation of variables
and spectral superposition.

To do so, we need a basis of eigenfunctions of the differential operator
$\D$ of equation~(\ref{ddef}).
Inspired by the explicit form of the solution~(\ref{fex}), we set
\be
G(u)=u^{\beta+1}\,\e^{-u^2/4}\,L(v),\qquad v=u^2/4.
\ee
The eigenvalue equation $(\D-E)G(u)=0$ becomes
\be
v\frac{\d^2 L}{\d v^2}
+\left(\frac{\beta+3}{2}-v\right)\frac{\d L}{\d v}+(E-1)L=0,
\ee
which is the differential equation
obeyed by the Laguerre polynomials $L_n^\alpha(v)$~\cite{bateman},
with $\alpha=(\beta+1)/2$ and $n=E-1$.

The eigenvalues of the operator $\D$ therefore read $E_n=n+1$,
with $n=0,1,\dots$
The associated eigenfunctions,
\be
G_n(u)=u^{\beta+1}\,\e^{-u^2/4}\,\L_n(u^2/4),
\ee
with~\cite{bateman}
\beq
\L_n(v)=\sum_{k=0}^n\frac{\Gamma\!\left(n+\frac{\beta+3}{2}\right)}
{\Gamma\!\left(k+\frac{\beta+3}{2}\right)}\frac{(-v)^k}{k!(n-k)!},
\label{lag}
\eeq
obey the orthogonality property
\beq
\int_0^\infty
G_m(u)\,G_n(u)\,u^{-\beta}\,\e^{u^2/4}\,\d u=N_n\,\delta_{m,n},
\qquad N_n=\frac{2^{\beta+2}\,\Gamma\!\left(n+\frac{\beta+3}{2}\right)}{n!}.
\label{orthal}
\eeq

The Mellin transform $M_{G_n}(z)$
of these eigenfunctions can again be evaluated in closed form:
\beq
M_{G_n}(z)=\frac
{2^{\beta-z}\,\Gamma\!\left(n+\frac{z+2}{2}\right)\,
\Gamma\!\left(\frac{\beta+1-z}{2}\right)}
{n!\,\Gamma\!\left(\frac{z+2}{2}\right)}
\qquad(\re z<\beta+1),
\label{melgn}
\eeq
where the normalization has been fixed by the condition~\cite{bateman}
$\L_n(0)=1$.

The method of spectral superposition consists in looking
for the solution $G(u,x)$ of equation~(\ref{gdif})
as a linear superposition of the form
\beq
G(u,x)=\sum_{n=0}^\infty a_n x^{-(n+3/2)} G_n(u).
\label{gexpand}
\eeq
The coefficients $a_n$ are determined by the initial condition~(\ref{initg}).
Using the orthogonality property~(\ref{orthal}), we obtain
\be
a_n=\frac{1}{N_n}\int_0^\infty u^2\,F(u)\,\L_n(u^2/4)\,\d u.
\ee
Then, using~(\ref{lag}) and~(\ref{melgn}), we are left with
\beq
a_n=\frac{2^{1-\beta}}{\Gamma\!\left(\frac{\beta+1}{2}\right)}
\,n!\sum_{k=0}^n\frac{(-)^k}{(2k+3)k!(n-k)!}
\frac{\Gamma\!\left(k+\frac{\beta+4}{2}\right)}
{\Gamma\!\left(k+\frac{\beta+3}{2}\right)}.
\label{an}
\eeq


\newpage
\begin{thebibliography}{99}

\bibitem{ehr} P. and T. Ehrenfest, Phys. Zeit. {\bf 8} (1907), 311.

\bibitem{ks} F. Kohlrausch and E. Schr\"odinger, Phys. Zeit. {\bf 27}
(1926), 306.

\bibitem{kac} M. Kac, Amer. Math. Monthly {\bf 54} (1947), 369.

\bibitem{kac2} M. Kac, in {\it Probability and Related Topics in Physical
Sciences}, Lectures in Applied Mathematics, vol.~{\bf 1~A} (American
Mathematical Society, 1959).

\bibitem{sieg} A.J.F. Siegert, Phys. Rev. {\bf 76} (1949), 1708.

\bibitem{hess} F.G. Hess, Amer. Math. Monthly {\bf 61} (1954), 323.

\bibitem{ritort} F. Ritort, Phys. Rev. Lett. {\bf 75} (1995), 1190.

\bibitem{fr} S. Franz and F. Ritort, Europhys. Lett. {\bf 31} (1995), 507;
J. Stat. Phys. {\bf 85} (1996), 131; J. Phys. A {\bf 30} (1997), L359.

\bibitem{gbm} C. Godr\`eche, J.P. Bouchaud, and M. M\'ezard, J. Phys. A
{\bf 28} (1995), L603.

\bibitem{gl} C. Godr\`eche and J.M. Luck, J. Phys. A {\bf 29} (1996), 1915;
J. Phys. A {\bf 30} (1997), 6245; J. Phys. A {\bf 32} (1999), 6033.

\bibitem{bia} P. Bialas, Z. Burda, and D. Johnston, Nucl. Phys. B {\bf 493}
(1997), 505; Nucl. Phys. B {\bf 542} (1999), 413; P. Bialas, L. Bogacz,
Z. Burda, and D. Johnston, Nucl. Phys. B {\bf 575} (2000), 599.

\bibitem{dgc} J.M. Drouffe, C. Godr\`eche, and F. Camia, J. Phys. A {\bf 31}
(1998), L19.

\bibitem{ck} L.F. Cugliandolo and J. Kurchan, J. Phys. A {\bf 27} (1994), 5749.

\bibitem{aging} For reviews, see: E. Vincent, J. Hammann, M. Ocio, J.P.
Bouchaud, and L.F. Cugliandolo, in {\it Complex Behavior of Glassy Systems},
Springer Lecture Notes in Physics {\bf 492} (1997), 184 (cond-mat/9607224);
J.P. Bouchaud, L.F. Cugliandolo, J. Kurchan, and M. M\'ezard,
in {\it Spin Glasses and Random Fields},
Directions in Condensed Matter Physics, vol.~{\bf 12},
ed. A.P. Young (World Scientific, Singapore, 1998) (cond-mat/9702070).

\bibitem{bray} A.J. Bray, Adv. Phys. {\bf 43} (1994), 357.

\bibitem{glglau} C. Godr\`eche and J.M. Luck, J. Phys. A {\bf 33} (2000), 1151.

\bibitem{glcrit} C. Godr\`eche and J.M. Luck, J. Phys. A {\bf 33} (2000), 9141.

\bibitem{heun} A. Ronveaux, {\it Heun's Differential Equations}
(Oxford University Press, Oxford, 1995).

\bibitem{kim} B.J. Kim, G.S. Jeon, and M.Y. Choi, Phys. Rev. Lett. {\bf 76}
(1996), 4648.

\bibitem{xzero} L.F. Cugliandolo, J. Kurchan, and L. Peliti, Phys. Rev. E
{\bf 55} (1997), 3898; A. Barrat, Phys. Rev. E {\bf 57} (1998), 3629;
L. Berthier, J.L. Barrat, and J. Kurchan, Eur. Phys. J. B {\bf 11} (1999), 635.

\bibitem{bateman} A. Erd\'elyi (ed.), {\it Higher Transcendental Functions}
(The Bateman Manuscript Project) (McGraw-Hill, New-York, 1953).

\end{thebibliography}
\end{document}